\documentclass[reprint,prl,amsmath,amssymb,floatfix]{revtex4-2}

\usepackage{color}
\usepackage{multirow}
\usepackage{bm}
\usepackage{psfrag}
\usepackage[latin1]{inputenc}
\usepackage[english]{babel}
\usepackage{amsfonts}
\usepackage{amsmath}
\usepackage{upgreek}
\usepackage[hidelinks]{hyperref}
\usepackage[dvipsnames]{xcolor}

\usepackage{pdfpages} 
\usepackage{pgffor} 

\makeatletter
\AtBeginDocument{\let\LS@rot\@undefined}
\makeatother

\def\supplementfilename{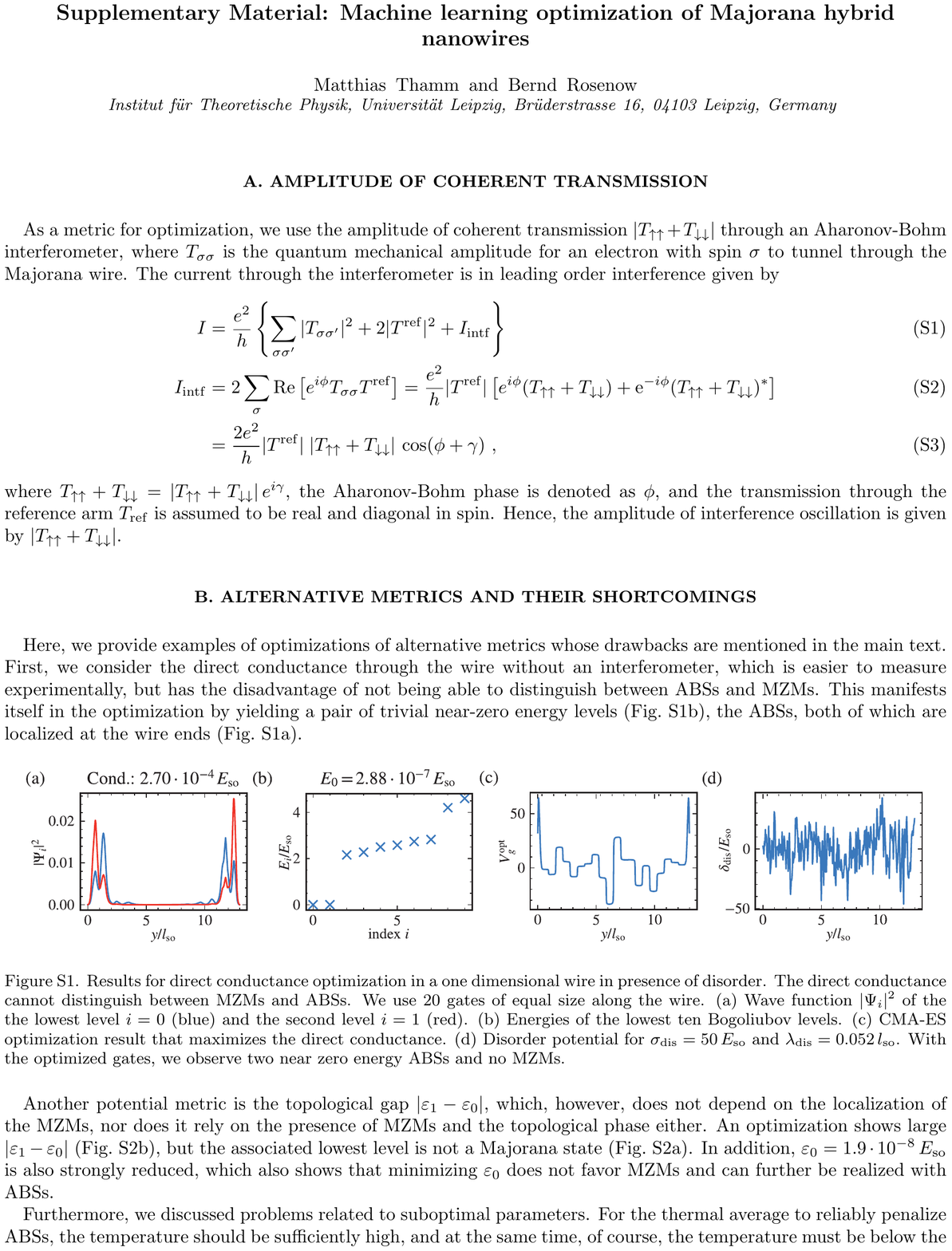}

\pdfximage{\supplementfilename}
\def\numbersupplementpages{\the\pdflastximagepages}

\newif\ifarXiv
\arXivtrue 

\DeclareSymbolFont{bbold}{U}{bbold}{m}{n}
\DeclareSymbolFontAlphabet{\mathbbold}{bbold}

\newcommand{\e}{{\rm e}}

\newcommand{\ex}[1]{\langle #1 \rangle}	
\newcommand{\up}{\uparrow}	
\newcommand{\down}{\downarrow}

\begin{document}

\title{
	Machine learning optimization of Majorana hybrid nanowires
	}
 
\author{
  Matthias Thamm  
	}  
\author{
	Bernd Rosenow
	}  
\affiliation{
	Institut f\"{u}r Theoretische Physik, Universit\"{a}t
  Leipzig,  Br\"{u}derstrasse 16, 04103 Leipzig, Germany
	} 
	
\date{\today}

\begin{abstract} 
		As the complexity of quantum systems such as quantum bit arrays increases,
		efforts to automate expensive tuning are increasingly worthwhile. We 
		investigate machine learning based tuning of gate arrays using the CMA-ES 
		algorithm for the case study of Majorana wires with strong disorder. We find 
		that the algorithm is able to efficiently improve the topological signatures, 
		learn intrinsic disorder profiles, and completely eliminate disorder effects.
		For example, with only 20 gates, it is possible to fully recover Majorana 
		zero modes destroyed by disorder by optimizing gate voltages.

\end{abstract}
 
\maketitle 
   
\emph{Introduction:}
	In recent years, increasingly complex quantum devices have been proposed and 
	implemented \cite{Karzig.2017,Arute.2019,Lennon.2019,Oreg.2020,Ares.2021}, 
	requiring more personnel-intensive tuning. Therefore, it is becoming 
	profitable, and in some cases even necessary, to automate the tuning process 
	\cite{Baart.2016,Craig.2021,Ziegler.2021}, and machine learning approaches have 
	been found to be very flexible and robust for this purpose \cite{Frees.2019,
	Kalantre.2019,Moon.2020,RuizEuler.2020,Ares.2021,Craig.2021,Ziegler.2021}. 
	Especially for the implementation of large scale quantum computation 
	\cite{Ladd.2010,Hanson.2007,Kloeffel.2013,Vandersypen.2017}, efficient tuning 
	of parameters and gates is crucial and numerous automations in quantum dot 
	based qubits have been proposed \cite{Baart.2016,Botzem.2018,Kalantre.2019,
	Teske.2019,Mills.2019,Durrer.2020,Moon.2020,vanEsbroeck.2020,Craig.2021,
	Fedele.2021,Ziegler.2021,Krause.2022}.

	A popular platform for scalable qubit architectures is based on Majorana zero 
	modes (MZMs) in topological superconductors \cite{Alicea.2011,Clarke.2011,
	Hyart.2013,Sarma.2015,Aasen.2016,Karzig.2017,Lutchyn.2018,Oreg.2020}, whose 
	advantages are the non-local storage of quantum information and its 
	manipulation via  anyonic braiding \cite{Alicea.2011,Clarke.2011,Hyart.2013,
	Sarma.2015,Vijay.2016}. MZMs have been proposed to exist in 
	semiconductor-superconductor heterostructures \cite{Kitaev.2001,Lutchyn.2010,
	Oreg.2010,Sau.2010,Alicea.2011} and many of their predicted signatures have 
	been observed, such as zero-bias conductance peaks \cite{Mourik.2012,Das.2012,
	Deng.2012,Nichele.2017}, the fractional Josephson effect \cite{Rokhinson.2012}, 
	and the suppression of even-odd splitting difference of conductance resonances
	in Coulomb blockade \cite{Albrecht.2016}. 
	For a clean wire, it has been theoretically demonstrated that a harmonic 
	potential profile \cite{Boutin.2018} and specially chosen magnetic field 
	textures \cite{Klinovaja.2012,Boutin.2018,Mohanta.2019,Turcotte.2020} can make MZMs more 
	robust, and the geometry of Majorana Josephson junctions has been optimized 
	to increase the size of the topological gap \cite{Melo.2022}. 
	Nevertheless, disorder remains a crucial problem \cite{Zhang.2017,
	Ahn.2021,DasSarma.2021,Yu.2021}  in such systems, as it can mimic MZM 
	signatures even in the topologically trivial region \cite{Bagrets.2012,
	Pikulin.2012,Liu.2012,Zhang.2017,Pan.2020,DasSarma.2021}, or destroy the 
	topological phase altogether \cite{Takei.2013}. 

		\begin{figure}[t!]
			\centering
			\includegraphics[width=8.6cm]{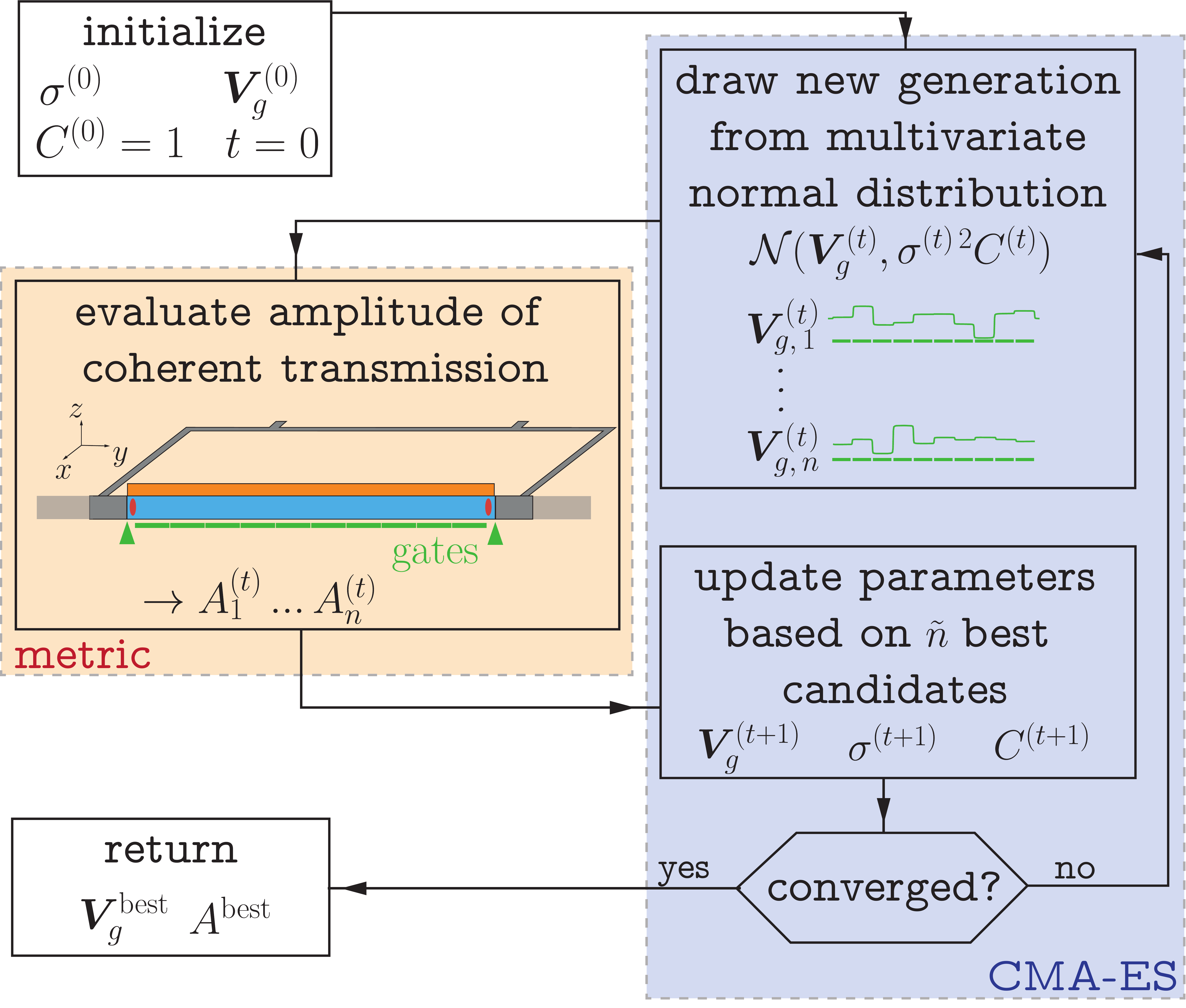}
			\caption{\label{Fig:sys}
							Schematic diagram of the Covariance Matrix Adaptation Evolution 
							Strategy (CMA-ES) algorithm \cite{Hansen.2001,Hansen.2003} 
							 used to learn an optimal gate voltage configuration that 
							maximizes the amplitude $A$ of coherent transmission through a 
							Majorana hybrid wire embedded into one arm of an Aharonov-Bohm 
							interferometer. Initially, one sets a step size $\sigma^{(0)}$, a
							covariance matrix $C^{(0)}$, and  starting gate voltages 
							$\bm{V}_g^{(0)}$. In each iteration $t$, $n$ gate voltage 
							configurations are drawn from a multivariate normal 
							distribution with mean $\bm{V}_g^{(t)}$ and covariance $C^{(t)}$.
							Based on  the amplitudes $A_i$ for the proposed gate voltage 
							configurations $\bm{V}_{g,i}^{(t)}$, the  new mean value 
							$\bm{V}_g^{(t+1)}$ is determined, and step size $\sigma^{(t+1)}$ 
							and covariance $C^{(t+1)}$ are updated.}
		\end{figure}

	In this letter,  we present a case study of automatic tuning of a gate array in 
	proximity to a strongly disordered Majorana wire  using the CMA-ES 
	\cite{Hansen.2001,Hansen.2003} algorithm. CMA-ES is a machine learning 
	algorithm that does not need  system specific information to operate, and is 
	widely applicable for high dimensional optimization problems 
	\cite{Hansen.2006,Lozano.2006,Loshchilov.2016,WilljuiceIruthayarajan.2010,
	Loshchilov.2013}.
   A crucial requirement is to find a good metric such that desired properties 
	of the physical system are indeed improved during optimization.  
  For example, signatures of MZMs can be mimicked by topologically trivial 
	Andreev bound states (ABSs) \cite{Kells.2012,Prada.2012,Rainis.2013, Cayao.2015,
	San.2016,Chen.2017,Liu.2017,Penaranda.2018,Avila.2019,Chiu.2019,Chen.2019,
	Woods.2019,Vuik.2019,Dmytruk.2020,Pan.2020,Prada.2020,Valentini.2021,Zhang.2021}, 
	which one would like to avoid. We therefore use the amplitude of coherent 
	transmission \cite{Fu.2010,Hell.2018,Whiticar.2020,Thamm.2021} through a 
	Coulomb-blocked Majorana wire as a metric, which can be measured by placing 
	the wire in an arm of an electron interferometer \cite{Whiticar.2020} and 
	allows to distinguish MZMs from ABSs \cite{Hell.2018,Thamm.2021}. 
	We find that 	as little as a few 100 to some 1000 amplitude measurements are 
	sufficient to tune the gate array, such that  (i)  both the localization of 
	the MZMs and the transmission amplitude are significantly improved, and (ii)  
	strong potential disorder is compensated. 
			
	\begin{figure}[tp]
			\centering
			\includegraphics[width=8.6cm]{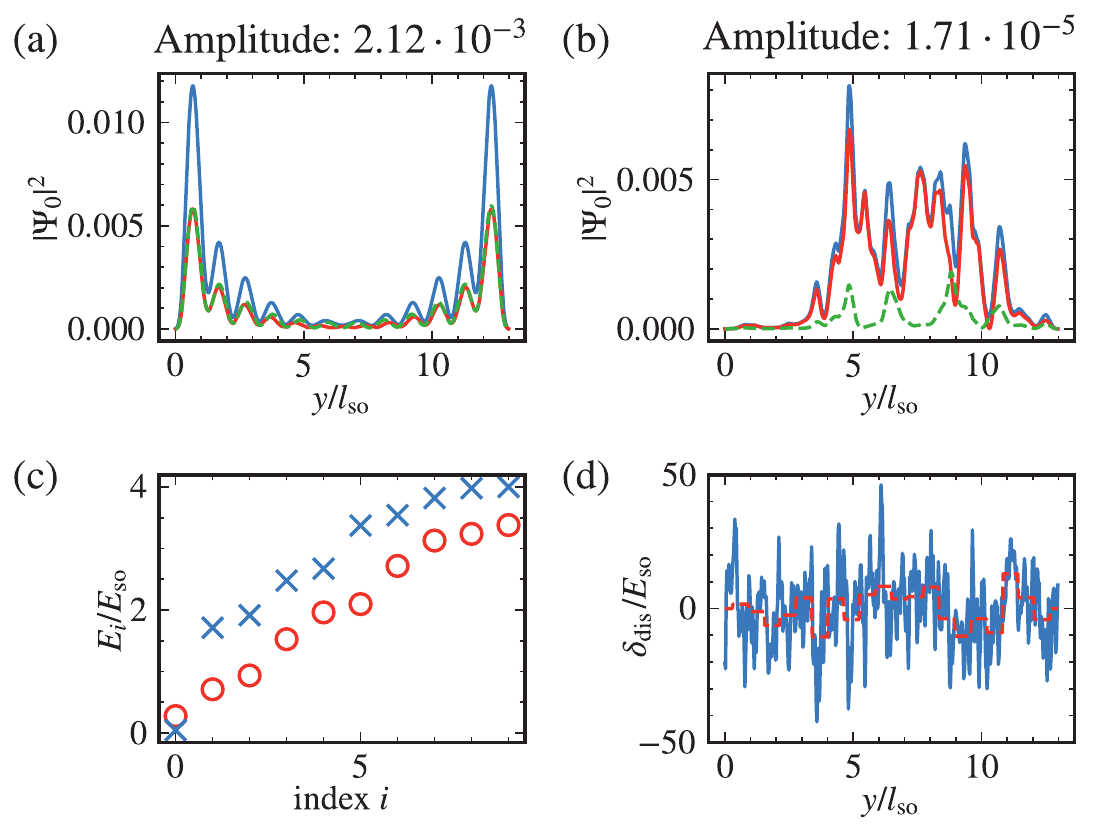}
			\caption{\label{Fig:refAndRefDiss} Reference results for deactivated
																	gates: wave function $|\Psi_0|^2$ of
																	the lowest level (blue), decomposed into 
																	hole wave function $|\bm{v}_0|^2$ (orange) and
																	electron wave function $|\bm{u}_0|^2$ (green)
																	for the case 
																	(a) without disorder,
																	(b) with disorder strength 
																	$\sigma_{\rm dis}=50\,E_{\rm so}$ and 
																	correlation length
																	$\lambda_{\rm dis}=0.052\,l_{\rm so}$.
																	(c) Energies of the lowest ten Bogoliubov 
															    levels for the case without disorder (blue 
																	crosses) and with disorder (red circles). 
																	(d) Disorder potential along the wire (blue) 
																	and average	of the disorder over  gate regions 
																	(dashed, red).
																	}
		\end{figure}

\emph{Setup:} 
	A schematic of the Majorana hybrid wire embedded in one arm of an 
	Aharonov-Bohm interferometer and a flow diagram of the CMA-ES algorithm are
	shown in Fig.~\ref{Fig:sys}. Our goal is to find voltages $V_j$ of $N_{\rm g}$ 
	gates such that the localization of MZMs inside the wires as well as their 
	tunnel coupling to the connecting wires are optimized.  For optimization 
	purposes we Fourier expand the gate voltages as
	\begin{align}
		{V}_j = \frac{b_0}{2} 
			&+ \sum_{k=1}^{\left\lfloor \frac{N_{\rm g}-1}{2}\right\rfloor}
			a_k \sin\left(\frac{2\pi}{N_{\rm g}}\, k j\right) \notag\\
			&+ \sum_{k=1}^{\left\lfloor \frac{N_{\rm g}}{2}\right\rfloor}
			b_k \cos\left(\frac{2\pi}{N_{\rm g}}\, k j\right)
			\ .
	\end{align}
	Optimization of Fourier components has the advantage that by choosing $b_0=0$ 
	and optimizing the remaining $N_{\rm g}-1$ components, the gate voltage is 
	zero on average, whereas direct optimization of all the $V_j$ and then 
	manually removing the average according to  $V_j \to {V}_j - \ex{V_j}$ 
	negatively effects the robustness of the algorithm.

	To obtain the spatial potential profile $V_g(y)$ acting on the hybrid wire, we
	assume that the wire is located at a distance $z_{0}$  from the gates, such 
	that the potential profile is smoothed according to 
	\begin{align}
		V_g(y) = \mathcal{F}^{-1}\left[\e^{-|q|z_{0}} 
			\mathcal{F}\left[\sum_{j=1}^{N_{\rm g}} {V}_j\chi_j(y)\right]
			\right] \ , \label{Eq:Vg}
	\end{align}
	where $\mathcal{F}$ and $\mathcal{F}^{-1}$ are Fourier transform and inverse 
	Fourier transform in the variables $y$ and $q$, respectively, and $\chi_j(y)$ 
	is the characteristic function of gate $j$, i.e., $\chi_j(y)$ is 1 if $y$ lies
	in the region of gate $j$ and 0 otherwise.

  We first consider a strictly one-dimension model for the hybrid wire, and 
	generalize to a more realistic two-dimensional model later. 
  The Majorana wire consisting of a semiconductor with Rashba spin-orbit 
	coupling $\alpha_R$ and a proximity induced  s-wave gap $\Delta$ is described 
	in the Nambu basis $\left(d_\up^\dagger(y),d_\down^\dagger(y),d_\down(y),
	-d_\up(y)\right)$ by the Hamiltonian
	\begin{align}
		\mathcal{H}_{\rm wire} = \tau_z \Bigg[ 
					&-\frac{\hbar^2\partial_y^2}{2m^*}\sigma_0 - \mu\sigma_0 
					-i\hbar\alpha_R\sigma_x\partial_y 
					\notag\\
					&+ \delta_{\rm dis}(y)\sigma_0 + V_g(y)\sigma_0+V_{\rm conf}(y) \sigma_0
					\Bigg] 
					\notag\\
					&-E_z\tau_0\sigma_z + \Delta\tau_x\sigma_0
					\ , \label{Eq:Hwire}
	\end{align}
	with disorder potential $\delta_{\rm dis}$, confinement potential 
	$V_{\rm conf}$, gate potential $V_g$ (see Eq.~\eqref{Eq:Vg}), and Pauli
	matrices $\sigma_i$ and $\tau_i$ acting in spin and particle-hole space, 
	respectively.
	Rashba spin-orbit coupling defines a characteristic energy scale 
	$E_{\rm so}=\alpha_R^2 m^*/2 = 0.05\,meV$ and length scale $l_{\rm so}
	=\hbar/(\alpha_R m^*)=0.19\,\mu m$ of the system, where $\hbar\alpha_R
	=0.2\,eV\rm\AA$ and $m^*=0.02\,m_e$ are realistic values for InAs 
	\cite{Mourik.2012,Lutchyn.2018}. 
	Throughout this paper, we consider wires of length $L=13\,l_{\rm so}$ on a 
	grid with spacing $a=0.026\,l_{\rm so}$. We use a
	chemical potential $\mu=1\,E_{\rm so}$, a Zeeman energy $E_z=6\,E_{\rm so}$,
	and  gap $\Delta=2\,E_{\rm so}$, such that the system
	in the absence of disorder and gate voltages is in the topological regime.
		
	We describe disorder in the wire by first drawing random numbers $\delta$ with 
	standard deviation $\sigma_{\rm dis}$ from a normal distribution and then 
	introduce a finite  correlation length $\lambda_{\rm dis}$ by damping 
	high Fourier modes according to 
	\begin{align}
		\delta_{\rm dis}(y) = \mathcal{F}^{-1}\left[
				\e^{-|q|\lambda_{\rm dis}} \mathcal{F}[\delta(y)]
				\right]\ .
	\end{align}
	Here the case $\lambda_{\rm dis}=0$ corresponds to onsite disorder.
	
	Wire and leads are connected via  steep tunnel barriers  
	of shape 
	$V_{\sigma,V_0}(y) = V_0 \exp[-y^2/(2\sigma^2)]$ with
	$\sigma=0.1\,l_{\rm so}$ and $V_0=65\,E_{\rm so}$ 
	which we assume to be defined by separate gates that are not included in the 
	optimization. For simplicity, we assume that the leads are normal conducting 
	and without spin orbit coupling. We treat Coulomb blockade in the Majorana 
	wire using a mean-field approximation, such that adding an electron to the 
	system of $N_0$ electrons costs an additional	charging energy 
	$E_c=8\,E_{\rm so}$, and introduce effective energy levels 
	$\varepsilon_{{\rm eff}, i}$ containing both charging energy and single 
	particle energies. We consider the system to be tuned to the center between 
	the conductance resonances for a fixed particle number $N_0$ in the Majorana
	wire.

	Finally, using the Weidenm\"{u}ller formula \cite{Mahaux.1968}
	\begin{align}
		\bm{T} = i\varphi_R^\dagger \Gamma_R U_w
					\frac{1}{\varepsilon - {\rm diag}(\varepsilon_{\rm eff}) 
					- U_w^\dagger \Sigma U_w}
					U_w^\dagger  \Gamma_L\varphi_L \ ,
	\end{align}
	with eigenvectors $U_w$ of the wire Hamiltonian Eq.~\eqref{Eq:Hwire}, we can 
	determine the transmission amplitude $A=|T_{\up\up}+T_{\down\down}|$ 
	in the middle between conductance resonances.
	Here, $\varepsilon$ is the energy of incoming electrons in the lead, and  
	self-energies $\Sigma_\alpha$, $\Gamma_\alpha=i(\Sigma_\alpha-\Sigma_\alpha^\dagger)$, 
	$\Sigma=\sum_\alpha \Sigma_\alpha$, and propagating modes $\varphi_\alpha$ of 
	lead $\alpha$ are obtained  by using the Python package KWANT \cite{Groth.2014}.

		\begin{figure}[tp]
			\centering
			\includegraphics[width=8.6cm]{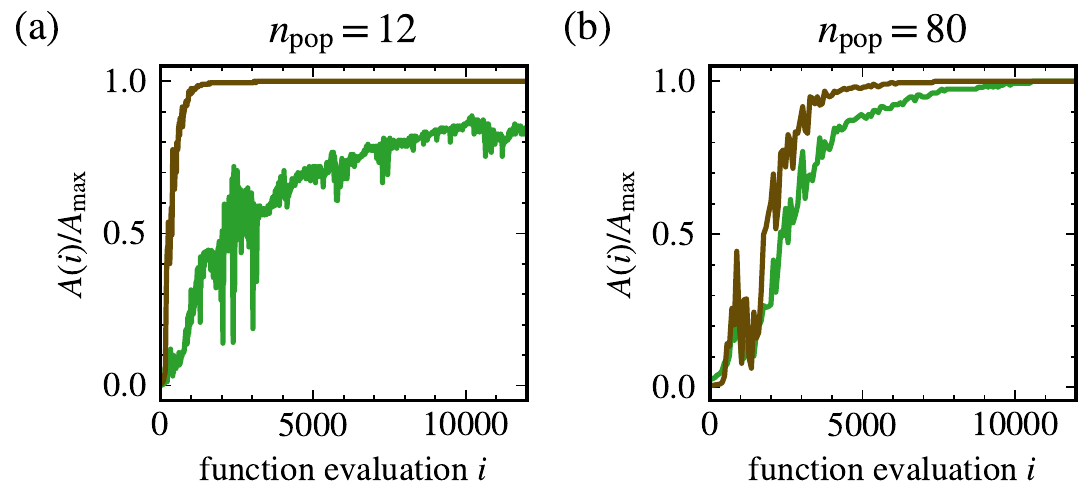}
			\caption{\label{Fig:convergence}Convergence speed of the CMA-ES 
																algorithm  in the presence of disorder using 
																20 gates.
																We consider runs of the CMA-ES algorithm with
																population sizes (a) $n_{\rm pop}=12$ and (b)
																$n_{\rm pop}=80$. 
																In both cases, we consider a ``hard'' problem 
																(green)	with onsite disorder, and an ``easier'' 
																problem with short range disorder correlations 
																$\lambda_{\rm dis} = 0.052\,l_{\rm so}$.
																The panels depict the degree of convergence 
																$A(i)/A_{\rm max}$ as a function of the
																number of function evaluations $i$ (number of 
																amplitude measurements) where $A_{\rm max}$ is the
																value to which the amplitude ultimately converges.
																	}
		\end{figure}
	A finite temperature can be considered by computing the scattering matrix for
	different thermal excitations of the wire and  thermally averaging the 
	transmission amplitude. Then, the transmission amplitude vanishes in the case
	of trivial ABSs \cite{Hell.2018,Thamm.2021}, such that our metric is able to 
	distinguish true MZM from an ABS. For the optimization, we consider transport 
	through the first 10 levels and verify the final results by taking into 
	account 50 levels. We carefully checked	that this does not influence the 
	optimization results.
	
	In the absence of disorder and with zero voltage at all gates, the lowest 
	level of the wire is approximately at zero energy in the middle of the 
	topological gap (see Fig.~\ref{Fig:refAndRefDiss}c), and the associated wave
	function $\Psi_0=(\bm{u}_0,\bm{v}_0)$ is localized at the wire ends and 
	satisfies the Majorana condition $|\bm{u}_0(y)|=|\bm{v}_0(y)|$ 
	(Fig.~\ref{Fig:refAndRefDiss}a). However, if one adds strong disorder 
	(Fig.~\ref{Fig:refAndRefDiss}d), both topological gap (red circles, 
	Fig.~\ref{Fig:refAndRefDiss}c) and MZMs (Fig.~\ref{Fig:refAndRefDiss}b) are 
	destroyed. As a result, the associated transmission amplitude is reduced by 
	two orders of magnitude as compared to the clean wire.

		\begin{figure}[tp]
			\centering
			\includegraphics[width=8.6cm]{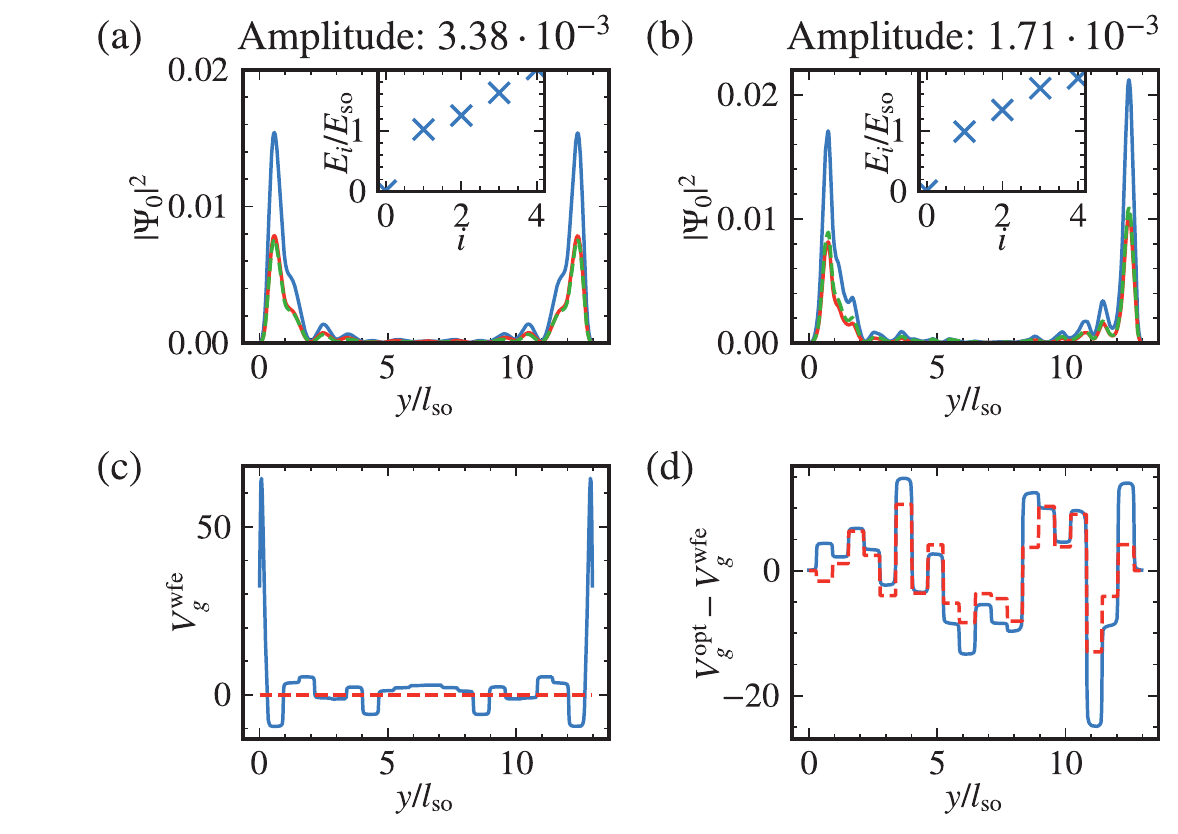}
			\caption{\label{Fig:wfeAndCDopt}Optimization of transmission via tuning of 
														 20 gates for a one-dimensional wire in the 
														 topological regime with $n_{\rm pop}=80$. 
														 Wave function $|\Psi_0|^2$ of the Majorana 
														 level (blue), and the corresponding hole and 
														 electron wave functions 
														 $|\bm{v}_0|^2$ (orange),  
														 $|\bm{u}_0|^2$ (green) for 
														 (a) no disorder in the wire and
														 (b)  disorder strength $\sigma_{\rm dis}
														 =50\,E_{\rm so}$ and a correlation length
														 $\lambda_{\rm dis}=0.052\,l_{\rm so}$.
														 The insets depict the energies of the lowest five 
														 Bogoliubov levels. Optimized gate potentials in the 
														 absence of disorder are shown in (c), and in (d) the 
														 difference between  optimized  potential obtained 
														 with and without disorder is shown. The dashed,
														 red line shows the negative average 
														 $-V_{\rm dis}^{\rm avg}$ of the disorder potential 
														 over the gates, indicating that the algorithm has
														 learned the shape of the disorder potential. 
														}
		\end{figure}

\emph{Optimization results:}
	To understand the convergence behavior and the influence of the population 
	size $n_{\rm pop}$ on the CMA-ES algorithm, we  consider two scenarios: 
	(i) disorder with a finite correlation length $\lambda_{\rm dis}
	=0.052\,l_{\rm so}$  and (ii) onsite disorder. For both cases, we perform a CMA
	-ES optimization of 20 gates  with population sizes $n_{\rm pop}=12$ and 
	$n_{\rm pop}=80$. In the easier case (i) already the smaller population size 
	is sufficient to achieve fast convergence after less than 1000 function 
	evaluations (brown line Fig.~\ref{Fig:convergence}a), whereas for  
	$n_{\rm pop}=80$  about five times as many evaluations	are necessary  (brown 
	line Fig.~\ref{Fig:convergence}b). In contrast, we find that the more 
	difficult problem (ii) converges poorly	in the case of small population sizes,
	but converges almost as fast as the correlated disorder case for 
	$n_{\rm pop}=80$.   
	Thus, if the primary time effort is to perform a function evaluation, i.e., 
	a measurement of the metric in the experiment,  we recommend to deviate from 
	the standard value $n_{\rm pop}=4+3\ln(N_{\rm g}-1)$ \cite{pycma.2019} for the
	case of a small disorder correlation length.  
	
	\begin{figure}[tp]
			\centering
			\includegraphics[width=8.6cm]{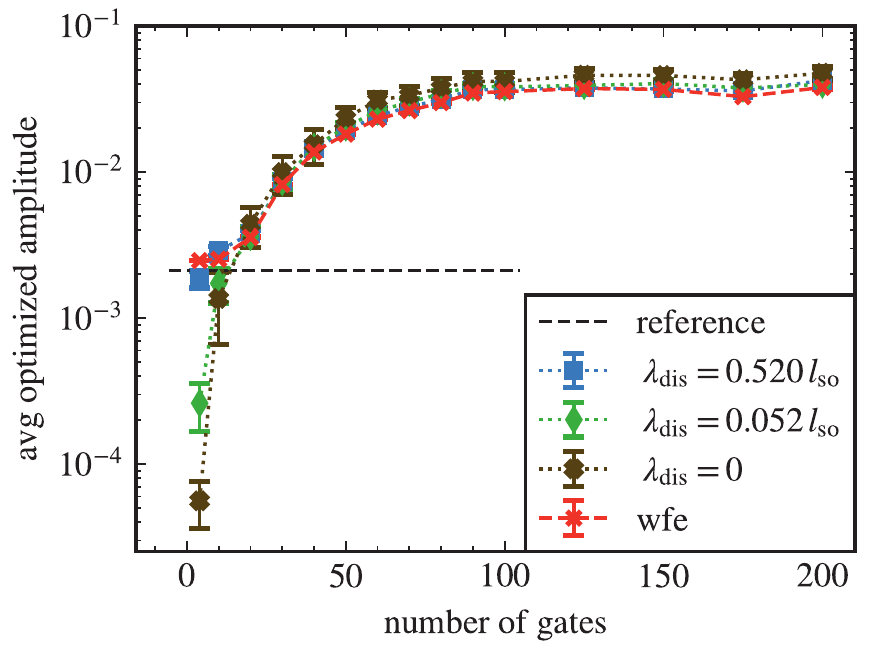}
			\caption{\label{Fig:ampOfNgates}Optimized transmission amplitude as a 
																			function of the	number of gates along the 
																			wire. Results are shown for disorder  
																			strength $\sigma_{\rm dis}=50\,E_{\rm so}$ 
																			and  correlation lengths 
																			$\lambda_{\rm dis}=0.52\,l_{\rm so}$ (blue
																			squares), $0.052\,l_{\rm so}$ (green 
																			diamonds), $0$ (onsite disorder, brown 
																			crosses). Red crosses indicate the wave 
																			function engineering result obtained for
																			optimization without any disorder in the 
																			wire.
																			We show averages over ten realizations of 
																			disorder in each case, and an average over 
																			10 seeds of the CMA-ES algorithm in the 
																			absence of disorder. The black dashed line 
																			shows the reference value obtained
																			without disorder and no optimization.
																			 }
		\end{figure}

	We distinguish between two different types of optimizations in the following: 
	(i)  optimization in the absence of disorder to determine what shape a 
	potential should have to improve the localization properties of the MZMs 
	("wave function engineering"), and (ii) optimization with disorder in the 
	wire.  
	In the case of wave function engineering for 20 gates, we find an enhancement
	of the transmission amplitude by a factor of about $1.6$ by optimizing  the
	localization of the MZMs (Fig.~\ref{Fig:wfeAndCDopt}a) while keeping a sizable
	topological gap (inset). To achieve this, potentials of the outermost gates 
	are lowered to draw more weight of the wave functions to the wire ends 
	(Fig.~\ref{Fig:wfeAndCDopt}c, \cite{Supplement}).
	In case (ii) with disorder (c.f.~Fig.~\ref{Fig:refAndRefDiss}b), the 
	optimization almost completely restores the MZMs and the topological gap, 
	increasing the transmission amplitude by two orders of magnitude 
	(Fig.~\ref{Fig:wfeAndCDopt}b). This is achieved by the optimized potential 
	compensating the  average disorder (dashed red line	Fig.~\ref{Fig:wfeAndCDopt}d),  
	in addition to the zero disorder optimal values (Fig.~\ref{Fig:wfeAndCDopt}d). 
	We emphasize  that the CMA-ES algorithm has no knowledge about system 
	parameters, but only suggests gate configurations based on corresponding 
	transmission amplitudes.

	\begin{figure}[tp]
			\centering
			\includegraphics[width=8.6cm]{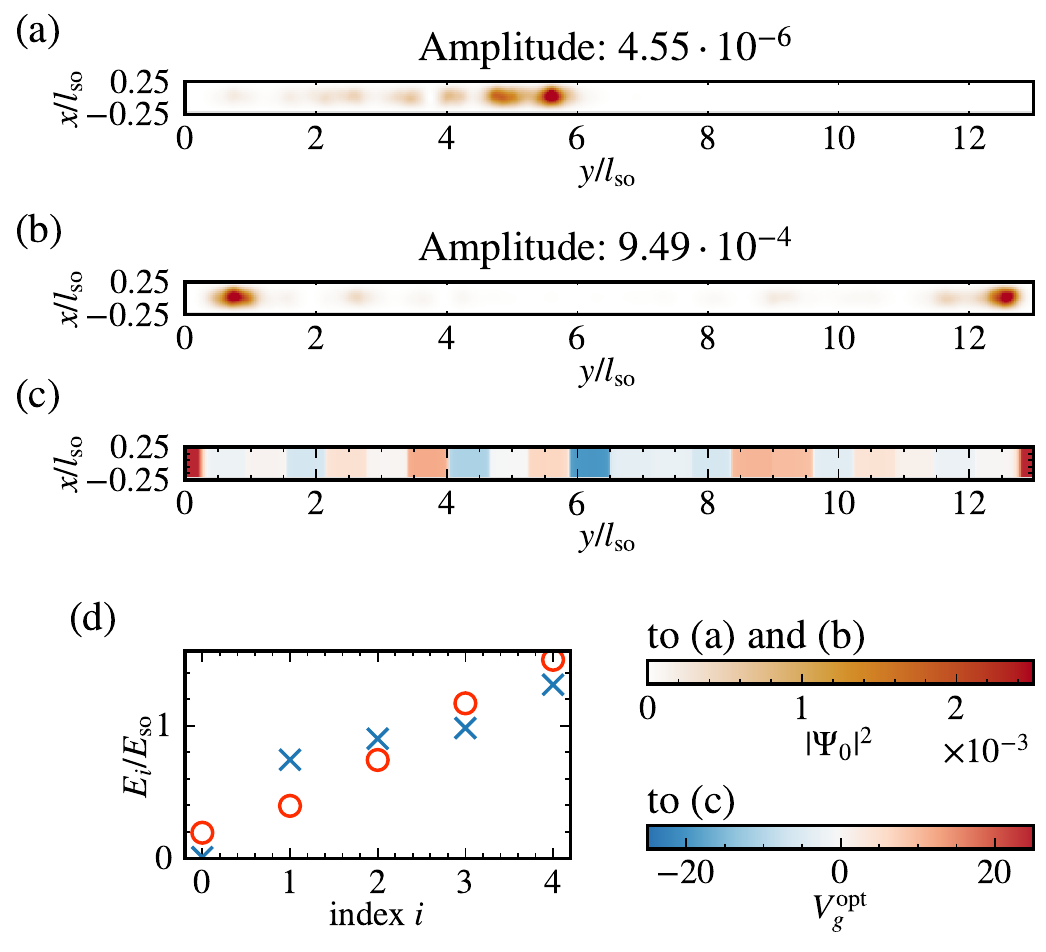}
			\caption{\label{Fig:2dopt}Optimization of 20 gates along a two-dimensional
														wire in the topological regime. 
														Wave function $|\Psi_0|^2$ of the lowest level for 
														(a) disorder with $\sigma_{\rm dis}=150\,E_{\rm so}$
														and $\lambda_{\rm dis}=0.052\,l_{\rm so}$ where all 
														gates are set to zero for reference and 
														(b) optimized gates in presence of disorder. 
														(c) CMA-ES optimization result for the gate potential
														that maximizes the transmission amplitude.
													  (d) Energies of the lowest ten Bogoliubov levels 
														for the reference case with disorder (red circles) 
														and for the optimized gate potential (blue crosses). 
														Similarly to the one dimensional case, the Majorana 
														zero modes, topological gap, and transmission 
														amplitude are restored by the	optimized gates.}
	\end{figure}

	Having seen from the examples how optimization can make MZMs more robust, we 
	next consider how reliable the optimization is for different disorder 
	correlation lengths, how many gates are necessary, and how strong the 
	dependence on the seed of the CMA-ES random number generator is. For this, 
	we consider 15 different values for the number of gates, from  $N_{\rm g}=4$ to 
	$N_{\rm g}=200$,  and three types of disorder, onsite ($\lambda_{\rm dis}=0$), 
	$\lambda_{\rm dis}=0.052\,l_{\rm so}$ and 
	$\lambda_{\rm dis}=0.52\,l_{\rm so}$, as well as wave function 
	engineering without disorder. 
	For the case with disorder, we consider ten different disorder realizations 
	and average the resulting amplitudes, while in the absence of disorder we 
	average over  ten different seeds of the CMA-ES algorithm. 
	We find that for at least	20 gates all considered disorder profiles can be 
	compensated reliably (see  Fig.~\ref{Fig:ampOfNgates}). For  too few gates 
	$N_{\rm g}\leq10$, it is no longer possible to remove disorder with very 
	small correlation length.   
	For many gates, $N_{\rm g} \approx 100$, the amplitude saturates, having 
	increased by one order of magnitude as compared to $N_{\rm g} = 20$, but with 
	the drawback that up to $10^5$ function evaluations are needed to achieve full 
	convergence. We observe a sweet spot $20 \leq N_{\rm g} \leq 50$, where the 
	number of necessary function evaluations is acceptable and still significant
	improvements of the amplitude and complete compensation of disorder are 
	possible.

\emph{Choice of metric:} 
	Above, we have chosen the coherent transmission amplitude, since it 
	distinguishes ABS from MZMs \cite{Hell.2018,Thamm.2021} and benefits from 
	enhanced localization of MZMs. For a Majorana wire, other metrics come to mind
	that may be easier to determine experimentally, which however turn out to 
	cause problems in the optimization process. 
	For example, optimizing the gap $|\varepsilon_1-\varepsilon_0|$ has the 
	disadvantage that it does not require $\varepsilon_0$ to be small and	in 
	addition does not depend on the localization at the wire ends. 
	On the other hand,	when minimizing the lowest level $\varepsilon_0$,  
	localization of MZMs is not strengthened, and in addition one is not able to 
	exclude a vanishing gap or the presence of ABSs. Optimizing the incoherent 
	part of the conductance through the wire   produces trivial ABSs instead
	of MZMs by lowering the outermost gates \cite{Supplement} to create potential 
	wells at the ends \cite{Liu.2017}, while increasing the effective chemical 
	potential in the remaining wire, thus lifting	it into the trivial regime. 

\emph{Two dimensional case:} 
	We study a wire with length $L_y=13\,l_{\rm so}$ and width 
	$L_x=0.39\,l_{\rm so}$,  and account for the orbital effect of the magnetic 
	field by adding Peierls phases $\e^{-i e/\hbar \int_{\bm{r}_1}^{\bm{r}_2} 
	\bm{A}\cdot d\bm{r}}$ to the hoppings from site $\bm{r}_1$ to site $\bm{r}_2$.   
	We choose a chemical potential 
	$\mu=63\,E_{\rm so}$ and Zeeman energy $E_z=6\,E_{\rm so}$ such that the wire 
	in the absence of disorder and gates is in the topological regime with one 
	occupied subband (a  discussion of transport through higher subbands can be 
	found in \cite{Supplement}).
	In the presence of strong disorder, the MZMs are destroyed (Fig. 6a) and the 
	gap collapses (red circles in Fig. 6d), but again optimization with only 20 
	gates along the wire can restore the MZMs (Fig. 6b) as well as the gap (blue 
	crosses in Fig. 6d) similar to the one dimensional case.

\emph{Conclusions:}
	We studied  machine learning optimization of a gate array using the 
	CMA-ES  algorithm.  Using the coherent transmission amplitude through a Coulomb 
	blockaded Majorana wire as metric, we find: (i) optimization in	absence of 
	disorder improves localization of MZMs significantly and (ii)	optimization even
	restores MZMs fully in the case of strong disorder that otherwise destroys the
	topological phase. We discussed the importance of the	choice of an appropriate
	metric, showed that the number of necessary function evaluations would be 
	experimentally feasible, and that a moderate number of gates is sufficient for
	restoration of MZMs in the presence of disorder.

\begin{acknowledgments}  
	\textit{Acknowledgments:} We would like to thank E.~van Nieuwenburg and A.~Chatterjee for helpful discussions. 
	 This work has been funded by the Deutsche Forschungsgemeinschaft (DFG) under
	 Grants No.~RO 2247/11-1 and No.~406116891 within the Research Training Group 
	 RTG 2522/1.\\ 
\end{acknowledgments}

%
	

\ifarXiv
    

\section*{Supplementary material}

# Supplementary Material: Machine learning optimization of Majorana hybrid nanowires

Matthias Thamm and Bernd Rosenow

*Institut für Theoretische Physik, Universität Leipzig, Brüderstrasse 16, 04103 Leipzig, Germany*

## A. AMPLITUDE OF COHERENT TRANSMISSION

As a metric for optimization, we use the amplitude of coherent transmission $|T_{\uparrow\uparrow} + T_{\downarrow\downarrow}|$ through an Aharonov-Bohm interferometer, where $T_{\sigma\sigma}$ is the quantum mechanical amplitude for an electron with spin $\sigma$ to tunnel through the Majorana wire. The current through the interferometer is in leading order interference given by

$$I = \frac{e^2}{h} \left\{ \sum_{\sigma\sigma'} |T_{\sigma\sigma'}|^2 + 2|T^{\text{ref}}|^2 + I_{\text{intf}} \right\} \tag{S1}$$

$$I_{\text{intf}} = 2 \sum_{\sigma} \text{Re} \left[ e^{i\phi} T_{\sigma\sigma} T^{\text{ref}} \right] = \frac{e^2}{h} |T^{\text{ref}}| \left[ e^{i\phi}(T_{\uparrow\uparrow} + T_{\downarrow\downarrow}) + e^{-i\phi}(T_{\uparrow\uparrow} + T_{\downarrow\downarrow})^* \right] \tag{S2}$$

$$= \frac{2e^2}{h} |T^{\text{ref}}| \, |T_{\uparrow\uparrow} + T_{\downarrow\downarrow}| \cos(\phi + \gamma) \, , \tag{S3}$$

where $T_{\uparrow\uparrow} + T_{\downarrow\downarrow} = |T_{\uparrow\uparrow} + T_{\downarrow\downarrow}| e^{i\gamma}$, the Aharonov-Bohm phase is denoted as $\phi$, and the transmission through the reference arm $T_{\text{ref}}$ is assumed to be real and diagonal in spin. Hence, the amplitude of interference oscillation is given by $|T_{\uparrow\uparrow} + T_{\downarrow\downarrow}|$.

## B. ALTERNATIVE METRICS AND THEIR SHORTCOMINGS

Here, we provide examples of optimizations of alternative metrics whose drawbacks are mentioned in the main text. First, we consider the direct conductance through the wire without an interferometer, which is easier to measure experimentally, but has the disadvantage of not being able to distinguish between ABSs and MZMs. This manifests itself in the optimization by yielding a pair of trivial near-zero energy levels (Fig. S1b), the ABSs, both of which are localized at the wire ends (Fig. S1a).

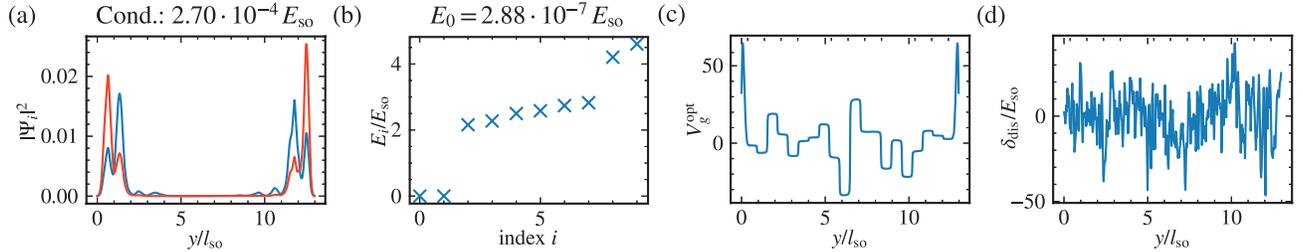

Figure S1. Results for direct conductance optimization in a one dimensional wire in presence of disorder. The direct conductance cannot distinguish between MZMs and ABSs. We use 20 gates of equal size along the wire. (a) Wave function $|\Psi_i|^2$ of the the lowest level $i = 0$ (blue) and the second level $i = 1$ (red). (b) Energies of the lowest ten Bogoliubov levels. (c) CMA-ES optimization result that maximizes the direct conductance. (d) Disorder potential for $\sigma_{\text{dis}} = 50\,E_{\text{so}}$ and $\lambda_{\text{dis}} = 0.052\,l_{\text{so}}$. With the optimized gates, we observe two near zero energy ABSs and no MZMs.

Another potential metric is the topological gap $|\varepsilon_1 - \varepsilon_0|$, which, however, does not depend on the localization of the MZMs, nor does it rely on the presence of MZMs and the topological phase either. An optimization shows large $|\varepsilon_1 - \varepsilon_0|$ (Fig. S2b), but the associated lowest level is not a Majorana state (Fig. S2a). In addition, $\varepsilon_0 = 1.9 \cdot 10^{-8}\,E_{\text{so}}$ is also strongly reduced, which also shows that minimizing $\varepsilon_0$ does not favor MZMs and can further be realized with ABSs.

Furthermore, we discussed problems related to suboptimal parameters. For the thermal average to reliably penalize ABSs, the temperature should be sufficiently high, and at the same time, of course, the temperature must be below the

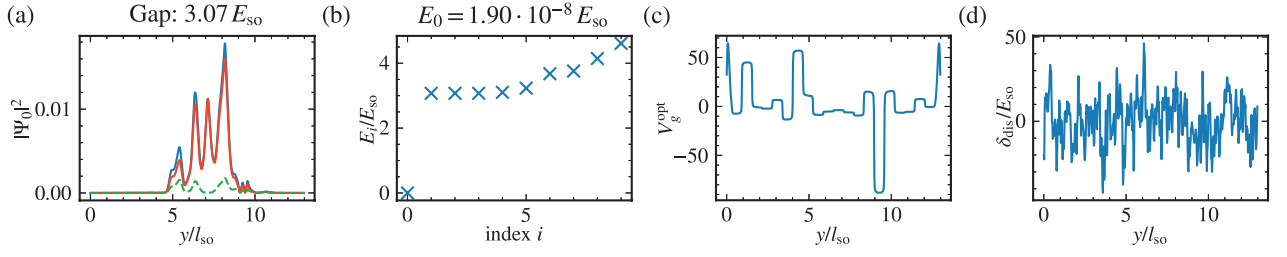

Figure S2. Results for gap optimization in a one dimensional wire in presence of disorder. We define the gap as the difference between the first two energy eigenvalues. We use 20 gates of equal size along the wire. (a) Wave function $|\Psi_0|^2$ of the the lowest level and corresponding hole and electron wave functions $|\boldsymbol{v}_0|^2$ (orange) and $|\boldsymbol{u}_0|^2$ (green). (b) Energies of the lowest ten Bogoliubov levels. (c) CMA-ES optimization result that maximizes the gap. (d) Disorder potential for $\sigma_{\mathrm{dis}} = 50\,E_{\mathrm{so}}$ and $\lambda_{\mathrm{dis}} = 0.052\,l_{\mathrm{so}}$. With the optimized gates, we observe an increased gap, however despite the lowest level being close to zero energy it is not a Majorana level as $|\boldsymbol{u}_0| \neq |\boldsymbol{v}_0|$.

critical temperature for preserving superconductivity. If one chooses too small temperatures, for example $T = 34\,\mathrm{mK}$, the optimization favors ABSs with energy slightly larger than temperature (Fig. S3) and also the gap above the ABS levels can be strongly reduced.

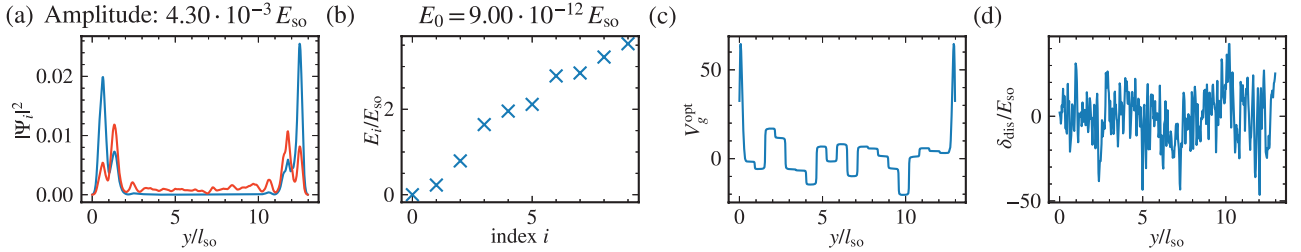

Figure S3. Results for transmission amplitude optimization in a one dimensional wire in presence of disorder but at very low temperature $\beta = 18\,E_{\mathrm{so}}^{-1}$. At very small temperatures an ABSs with low but finite energy has very different weight than a zero energy ABSs in the thermal average, such that the amplitude does not cancel when there are two ABSs with slightly split energy. We use 20 gates of equal size along the wire. (a) Wave function $|\Psi_i|^2$ of the the lowest level $i = 0$ (blue) and the second level $i = 1$ (red). (b) Energies of the lowest ten Bogoliubov levels. (c) CMA-ES optimization result that maximizes the gap. (d) Disorder potential for $\sigma_{\mathrm{dis}} = 50\,E_{\mathrm{so}}$ and $\lambda_{\mathrm{dis}} = 0.052\,l_{\mathrm{so}}$. With the optimized gates, we observe a diminished gap and a pair of ABSs near zero energy which are split by a small energy difference.

## C. COMPUTATIONAL DETAILS

We use the pycma [2] python implementation of the CMA-ES [3, 4] algorithm, with an initial configuration $\boldsymbol{V}_g^{(0)} = 0$, the starting step size $\sigma^{(0)} = 0.1\,E_{\mathrm{so}}$, population sizes of 80 or $4 + 3\ln(N_g)$, and a seed of the pseudo random number generator of 12345678, if not specified otherwise. As algorithm termination conditions, we use $tolfun = 10^{-15}$, $tolfunhist = 10^{-8}$, and $tolx = 10^{-5}\,E_{\mathrm{so}}$. We note, however, that the potentials do not change significantly anymore much earlier to meeting these conditions, such that one can stop the optimization earlier in an experimental situation.

For the computation of the transmission amplitude, we use KWANT [1] to obtain several quantities. We define the lattice, onsite terms, and hopping terms using KWANT (Fig. S4) and extract the self energies $\Sigma_\alpha$ of lead $\alpha$, as well as the propagating modes $\phi_\alpha$ defined at the lead-wire interfaces. In addition, KWANT allows to extract the wire Hamiltonian $\mathcal{H}_{\mathrm{wire}}$, which we use to compute eigenstates $U_w$ and energy levels $\varepsilon_w$, from which we obtain the effective couplings and energy levels [5] used in the computation of the scattering matrix. For the Majorana wires, we consider an effective mass $m^* = 0.02\,m_e$, Rashba spin orbit coupling strength $\hbar\alpha_R = 0.2\,\mathrm{eV\mathring{A}}$, a lattice spacing $a = 0.026\,l_{\mathrm{so}}$, and wire length $L = 13\,l_{\mathrm{so}}$. In addition, we set the chemical potential $\mu = 1\,E_{\mathrm{so}}$, the Zeeman energy $E_z = 6\,E_{\mathrm{so}}$, the proximity s-wave gap $\Delta = 2\,E_{\mathrm{so}}$, the charging energy $E_c = 8\,E_{\mathrm{so}}$, and the electron temperature $T = 183\,\mathrm{mK}$. In the wire, we use a steep confinement with $\sigma = 0.1\,l_{\mathrm{so}}$ and $V_0 = 65\,E_{\mathrm{so}}$ defined as $V_{\mathrm{conf}}(y) = V_{\sigma,V_0}(y - x_0) + V_{\sigma,V_0}(y - L + x_0)$ such that the maxima are located close to the ends of the wire at $x_0$ and $L - x_0$ where $x_0$ is chosen such that the

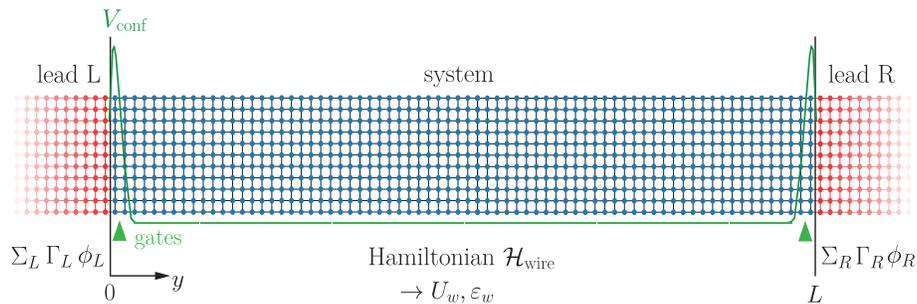

Figure S4. Sketch of the Majorana wire model. We consider leads (red) and wire (blue) of length $L$ to be separated by a confinement potential $V_{\text{conf}}$ (green). Using the python package KWANT [1], we define the lattice Hamiltonian, extract the lead self energies $\Sigma_\alpha$, $\Gamma_\alpha$, propagating modes $\phi_\alpha$, and the wire Hamiltonian matrix $\mathcal{H}_{\text{wire}}$ from which we obtain the eigenstates $U_w$ and energy levels $\varepsilon_w$.

potential has decayed to $V_0/2$ at the ends of the wire.

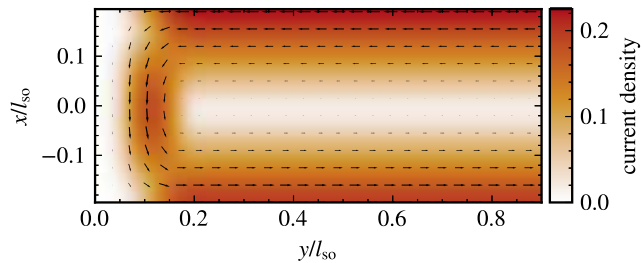

Figure S5. Supercurrent distribution $\boldsymbol{j}_S \propto \frac{\hbar}{m}\nabla\theta + \frac{2e}{m}\boldsymbol{A}$ at the left end of the two dimensional wire for the choice of vector potential $\boldsymbol{A}$ and superconducting order parameter phase $\theta$.

The potential in the leads is given by $V_{\text{lead}} = -100\,E_{\text{so}}$ to ensure that both spin components are present at the Fermi level. Leads are modeled with the Hamiltonian

$$\mathcal{H}_{\text{lead}} = \tau_z\left[-\frac{\hbar^2\partial_y^2}{2m^*}\sigma_0 + V_{\text{lead}}\sigma_0\right] - E_z\tau_0\sigma_z \ . \tag{S4}$$

We consider gates of equal extension that start a distance $0.3\,l_{\text{so}}$ from the ends of the wire to not interfere with the confinement potential which is produced by additional gates. Furthermore, we assume that the wire lies a distance $z_{\text{sys}} = 0.3\,l_{\text{so}}$ above the gates (see main text Eq. (2)).

## D. TWO-DIMENSIONAL WIRE

For the two dimensional case, we additionally choose a wire width $L_x = 0.39\,l_{\text{so}}$, chemical potential $\mu = 63\,E_{\text{so}}$, Zeeman energy $E_z = 6\,E_{\text{so}}$, and otherwise the same parameters as in the one dimensional case. The full 2d Hamiltonian is given by

$$\mathcal{H}_{\text{wire}}^{2d} = \tau_z\left[-\frac{\hbar^2}{2m^*}(\partial_x^2+\partial_y^2)\sigma_0 - \mu\sigma_0 - i\hbar\alpha_R(\sigma_x\partial_y - \sigma_y\partial_x) + \delta_{\text{dis}}(x,y)\sigma_0 + V_g(x,y)\sigma_0 + V_{\text{conf}}(y)\sigma_0\right]$$
$$+ \frac{\mu_B g B_z}{2}\tau_0\sigma_z + \Delta\tau_x\sigma_0 \ , \tag{S5}$$

with Lande factor $g = -14.9$ [6], and we take into account the orbital effect of the magnetic field by adding a Peierls phase $e^{-ie/\hbar\int_{r_1}^{r_2}\boldsymbol{A}\cdot d\boldsymbol{r}}$ to the hoppings. We choose the phase of the superconducting order parameter as $\theta = 0$ and, away from the wire ends, the vector potential as $\boldsymbol{A} = -B_z x\boldsymbol{e}_y$, so that it is independent of the coordinate $y$ along

the wire and the energy due to the supercurrent $\boldsymbol{j}_s = -2en_s(\hbar\nabla\theta + 2e\boldsymbol{A})/m$ is minimized [7]. At the wire ends we use the following approximation to guarantee current conservation:

$$\boldsymbol{A} = -a(y)B_z x \boldsymbol{e}_y + \frac{a'(y)}{2}B_z(x^2 - (L_x/2)^2)\boldsymbol{e}_x \ , \tag{S6}$$

$$a(y) = \begin{cases} f_{y_L, y_L + \lambda}(y) & y_L \le y > y_L + \lambda \\ 1 & y_L + \lambda \le y \le y_R - \lambda \\ 1 - f_{x_R - \lambda, x_R}(y) & y_R - \lambda < y \le y_R \end{cases} \ , \tag{S7}$$

$$f_{y_1, y_2}(y) = \frac{h(y - y_1)}{h(y - y_1) + h(y_2 - y)} \tag{S8}$$

$$h(y) = \begin{cases} \exp(-\lambda/y) & y > 0 \\ 0 & y \le 0 \end{cases} \ , \tag{S9}$$

which ensures that the vector potential at both ends $(x_L, x_R)$ of the wire vanishes over a distance $\lambda = L_x/2$ in a smooth manner. The resulting current $\boldsymbol{j}_s$ is shown in Fig. S5.

For computing the amplitude during optimization, we take into account the first ten effective energy levels, which speeds up the computations considerably, without influencing the transmission amplitude by a significant amount [5]. We verified this by evaluating the final transmission amplitude after optimization by taking into account 50 levels. In addition, we validated the amplitude in several cases at different steps during the optimization by considering all effective levels for single electron co-tunneling. We find that considering only ten levels during optimization adequately approximates taking the full number of levels into account, as it is relevant for an experiment.

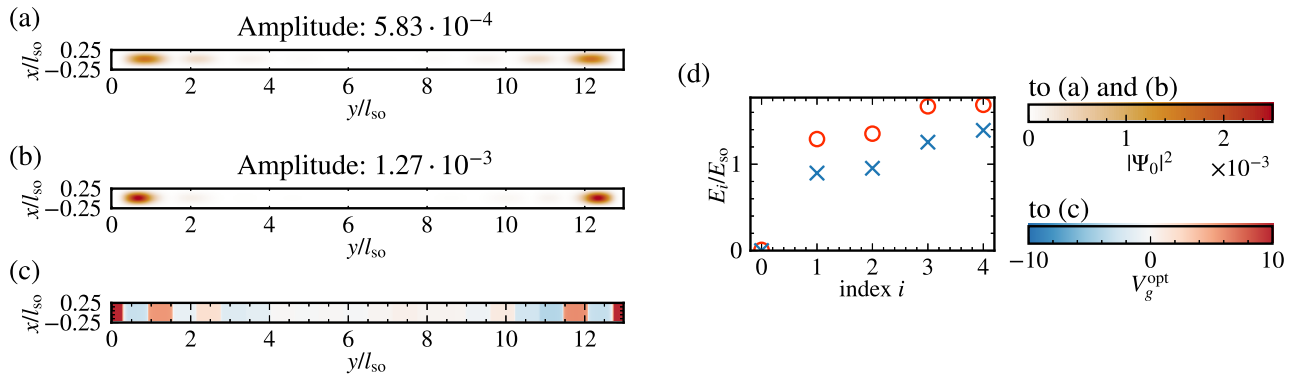

Figure S6. Results for transmission through a two dimensional wire in the topological regime for optimization of the gates. We use 20 gates of equal size along the wire. Wave function $|\Psi_0|^2$ of the lowest level for (a) the reference case without disorder and with zero gate voltage on all gates and (b) optimized gates without disorder (wave function engineering). (c) CMA-ES optimization result for the gate potential that maximizes the transmission amplitude. (d) Energies of the lowest five Bogoliubov levels for the reference case (red circles) and for the optimized gate potential (blue crosses).

## E. OPTIMIZATION IN THE SECOND SUBBAND OF THE TWO-DIMENSIONAL WIRE

Gate optimization can also be fruitful for higher subbands, as we show in Fig. S7 where we consider the second topological phase for $\mu = 144.5\,E_{so}$, $E_z = 6\,E_{so}$, $\Delta = 2\,E_{so}$. However, in the presence of levels from different subbands near the Fermi level many subtleties arise that can distract the CMA-ES optimization, such that the optimization is not always able to restore MZMs in presence of disorder. Importantly, different subbands have very different coupling strengths to the leads, e.g. MZMs in the second subband might have smaller couplings than topologically trivial states from the first subband [8]. In order to mitigate this effect, we move both the superconductor and the first/last gate a distance $1.04\,l_{so}$ away from the ends of the wire, and add on-site disorder with strength $\delta_{dis} = 100\,E_{so}$ to the superconductor-free region [8]. Only moving the superconductor away from the ends and having gates in the normal-conducting regions at the ends would allow the effective chemical potential in the superconductor to change such that the optimization is less stable with the risk of moving completely out of the topological regime. Using

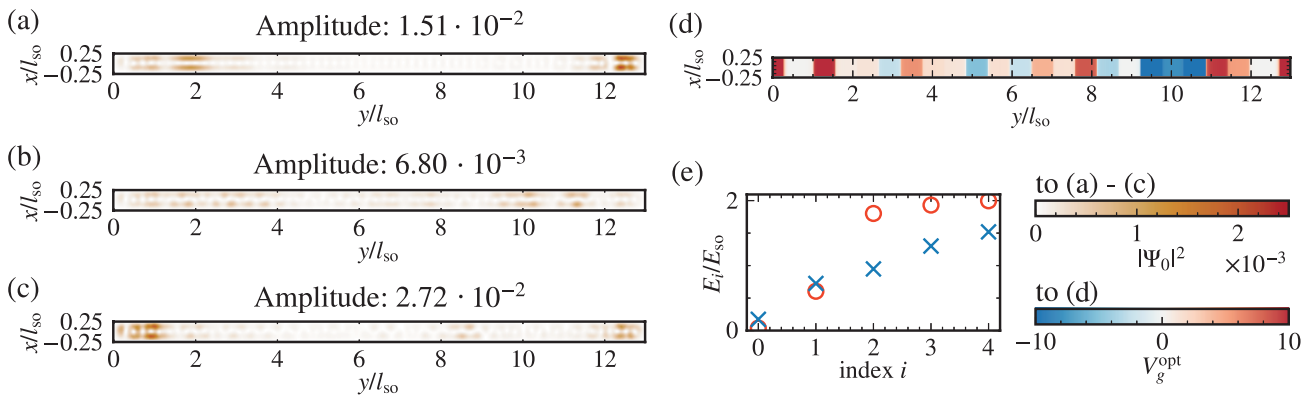

Figure S7. Results for transmission through a two dimensional wire in the topological regime in the second topological phase. Gates and superconductor are moved a distance $1.04\,l_{\mathrm{so}}$ away from the wire ends and onsite disorder with strength $100\,E_{\mathrm{so}}$ is added in the normal-conducting region to improve coupling of MZMs to the leads. We use 20 gates of equal size along the wire with $\mu = 244.5\,E_{\mathrm{so}}$, $E_z = 6\,E_{\mathrm{so}}$, and $\Delta = 2\,E_{\mathrm{so}}$. Wave function $|\Psi_0|^2$ of the lowest level for (a) the reference case without disorder and with zero gate voltage on all gates, (b) with bulk disorder ($\lambda_{\mathrm{dis}} = 0.052\,l_{\mathrm{so}}$, $\delta_{\mathrm{dis}} = 90\,E_{\mathrm{so}}$) before gate voltage optimization, and (c) with optimized gate voltages in the presence of disorder. (d) CMA-ES optimization result for the gate potential that maximizes the transmission amplitude. (e) Energies of the lowest five Bogoliubov levels for the reference case (red circles) and for the optimized gate potential (blue crosses).

the modified setup, we find MZMs, which in the reference case without bulk disorder (Fig. S7a) couple about one order of magnitude stronger to the leads then other low energy levels. When adding bulk disorder (Fig. S7b), they are destroyed and low energy levels couple with similar strength to the leads, and after optimization (Fig. S7c), the MZMs are restored with a coupling about twice as strong as other low energy levels. Even with these modifications, in presence of higher subbands at the Fermi level, the occurrence of Andreev bound states and other strongly coupling non-topological low energy states cannot reliably be excluded making the optimization overall more fragile. On the other hand, this also shows that CMA-ES optimization helps with identifying weaknesses in a given setup, such that it can also be used as a tool to test ways to stabilize desired features in the system.

---


\fi

\end{document}